# Social Play Spaces for Active Community Engagement

| Jenna Gavin | Ben Kenobi | Andy M. Connor |
|---|---|---|
| Auckland University of Technology | Auckland University of Technology | Auckland University of Technology |
| Private Bag 92006 | Private Bag 92006 | Private Bag 92006 |
| Wellesley Street, Auckland, NZ. | Wellesley Street, Auckland, NZ. | Wellesley Street, Auckland, NZ. |
| +64 (0)9 921 9999 | +64 (0)9 921 9999 | +64 (0)9 921 9999 |
| jengav14@aut.ac.nz | ben.kenobi@aut.ac.nz | andrew.connor@aut.ac.nz |

## ABSTRACT
This paper puts forward the perspective that social play spaces are opportunities to utilise both technology and body for the benefit of community culture and engagement. Co-located social gaming coupled with tangible interfaces offer active participant engagement and the development of the local video game scene. This paper includes a descriptive account of Rabble Room Arcade, an experimental social event combining custom-built physical interface devices and multiplayer video games.

## Categories and Subject Descriptors
H.5.2 [**User Interfaces**]: Input devices and strategies

## 1. INTRODUCTION
The video game industry has continued to grow dramatically over the past decade, cutting into mainstream media in participation and revenues as it becomes part of mainstream media culture [1]. Whilst gaming is sometimes (and naïvely) viewed by the public as an isolating activity, it is surprisingly social [2]. However, that social element is often related to collocated gameplay [2] rather than true social play. Social play is often characterised by play in pre-school children but this begs the question of why such play is not actively encouraged in older children, adolescents and adults? This paper outlines the design of a social play event that is based around challenging common perceptions related to video games and taking the concepts of gaming from collocated console play to one of physical, cooperative social play.

## 2. BACKGROUND & RELATED WORK

### 2.1 The Nature of Video Game Play
Video games have become immensely popular since popularised by the emergence of Pong in 1972. In today's society, video games are not just played on computers and game consoles but also handheld devices and cell phones. Because of the ubiquitous nature of these devices, games are no longer just played at home and at arcades, but are also played at work, at school, on public transport, and virtually anywhere that an electronic device can be operated.

The amount of time spent playing games has increased over time [3] and it is considered normal that children and adolescents play more than 20 hours each week with 40 hours of gaming not being uncommon among young males [4], with it being observed that so called "pathological gamers" spent twice as much time playing as nonpathological gamers and received poorer grades in school as well as exhibiting attention problems [5]. The almost obsessional growth in gaming has driven considerable research which has examined potential positive and negative effects of playing various types of video games. Much of this work has focused on the detrimental effects of playing violent games [6] or further exploring the negative association between time spent playing games and school performance [7]. However, gaming does present a particular dilemma as there is much research that emphasises the positive value associated with educational games [8], that games do have the potential to increase prosocial behaviour [9, 10] and that exercise games are an attractive form of physical activity [11, 12]. It seems that the impact of gaming depends on the game, the nature of play and the play environment.

### 2.2 The History of Video Gaming
The history of video gaming has been described in detail by many authors [13] so will not be considered in detail in this paper. However, reflection on the rise and fall of the industry provides insights in to the nature of play. Williams argues that the early 1980s were a crucial turning point in the social history of video game play that saw an erosion of what began as an open and free space for cultural and social mixing [14]. The history of video gaming can be summarised as slow adoption during the 1970s leading to a massive spike in popularity during the Atari heyday of the early 1980s, followed by the collapse of that company and the industry's eventual revival in the late 1980s by Nintendo.

The arcade establishment was the primary medium for the video game experience during the 1970's and 1980's, the golden age of arcade video games [15]. Despite the attention mandated by the video game screen, early arcade games were a carnival of physical experience, such as the 1975 eight-player game Indie 800, "which had a steering wheel and two pedals for each player" [15]. The video arcade machines were an offshoot from earlier mechanical games, such as pinball, and designers were attentive to the tangible interaction aspects of the game.

The late 1980s also saw the beginning of play moving from public to private spaces. Throughout the 1980s, a combination of economic and technological forces moved play away from social, communal and relatively anarchic early arcade spaces, and into the controlled environments of the sanitized mall arcade (or "family fun center") or into the home [14]. This was in part driven by the uptake of the home computer and game console in the 1990's, which shed much of the social and physical aspects of gaming. On this point, Salen and Zimmerman [16] describe single-player gaming as an anomaly in the rich history of games.

## 2.3 Motivation

While many modern video games embrace multiplayer modes through computer networking, screen-based gaming with a standardised button interface continues today as the main adult experience of games [17]. It has been argued that the input button so central to video gaming is impeding the development of the medium [17] because the button "[disregards] the bodies abilities" and permits the player to "forget about the physical device". The player is corporeally passive and detached. Researchers have issued a call to arms to abandon the button as soon as possible and replace it with more natural interfaces [18]. This paper argues that the button itself is not impeding the medium's development, but the conventions of usage surrounding the button. Designers and players have utilised technology to provide the cheapest and most efficient route to gratification. The social and physical implications include less physical exertion and less face-to-face bonding with others, whereas studies investigating more physical interfaces show the opposite [19].

This paper argues that there is a benefit in terms of more social and physically play, and proposes a return of the somewhat anarchical arcades of the 1970s and 1980s. The stereotype of gamers as antisocial creatures needs to be challenged, the emergence of LAN parties [20] suggests that gamers are in a way more tribal than solitary and thrive on the social aspects of play. Some authors go as far as to suggest that gaming is often as much about social interaction, as it is about interaction with the game content [21]. In the context of socially situated play, this paper argues that there is a place for community based play events that embrace the physicality of play as a means of increasing engagement and promoting the development of gaming media.

This view is borne out by the emergence of other local movements countering the potential isolation and sterilisation created by the use of modern technologies. For example, the New York collective Babycastles (Created by Kunal Gupta and Syed Salahuddin in 2010) provides a local play space to showcase artistic, independently-created video games and interfaces, alongside visual artists, installation artists and musicians. Similar projects are springing up around the world, including the LA Game Space, an inclusive workshop and gallery for people to explore the radical possibility of games.

In New Zealand, Guerilla Playspaces is an Auckland-based project that encourages public play through artifacts and installations. As Pasternack affirms, "the patrons of... these independent communities, are, in one way or another, striving to experience something new; something that can't be bought in a store, but that is available for anyone to see and hear if they look in the right places. Just like indie music, the independent gaming scene is trading in neat, mass-produced convenience for a rough-hewn, playful provocation" [22].

## 3. RABBLE ROOM ARCADE

Rabble Room Arcade was a project conceived and conducted by two undergraduate students studying for the Bachelor of Creative Technologies degree at the Auckland University of Technology. Rabble Room Arcade was the identity created for the social play event held in October 2013, and the choice of the word 'rabble', meaning the common people; disorderly crowd; or a "boisterous throng of people" intentionally focussed the context on community and agitation. The students set out to showcase independently-developed games, embracing the absurd, silly, and overtly physical, for the purpose of exciting a local cultural experience. Jane McGonigal [23] advocates gaming for change, especially in the face of global problems, arguing that "games are a sustainable way of life". When playing games with others, we ease our suffering, conserve resources, and participate in supportive and coordinated communities. Actively engaging with information can change values in relation to culture and potentially any number of topics. Active participation "increases the likelihood that one will learn from the video game due to greater identification and immersion" [24]. One form of active participation is when the *body is engaged* in play. Tangible interfaces allow "physically engaging experiences with technology" [25]. Wilson [26] affirms that "the material and social circumstances behind gameplay... play a key role in shaping any gameplay experience".

### 3.1 Inefficient Interfaces

A technique employed by Rabble Room Arcade was to explore interfaces that opposed optimised efficiency. This was done not only for the purpose of disrupting expectation (and thus encouraging active, divergent thought), but also to even the "playing field" as it were, so that the games did not privilege those that have trained for hours on standard interface devices. This "gestural excessiveness, as a showy form of inefficient gameplay, represents a refutation of hardcore instrumental play" [26]. The technique of designing for inefficiency works very well for social spectacle but may degrade with repeat and long-term play. This paper proposes that inefficiency of the interface and interactions should be considered in a light-hearted and social environment; for "especially in regards to party and street games, public spectacle comprises the heart and soul of what those activities are" [26].

### 3.2 Featured Games

Rabble Room Arcade featured eight very different games, including:

*Shadow Showdown* by Matthew Martin, Jenna Gavin, and Daniel Cermak-Sassenrath: A cooperative game where one or more players have to match silhouettes on the screen by creating silhouettes with their own body/bodies.

*Elevator* by CyrilQ Studios: A two player competitive game with cranks as input devices that have to be operated as fast as possible to make the game character go up an elevator as fast as possible while avoiding virtual objects being thrown at them.

*Double Shovel* by Jeff Nusz: A game where two players would cooperatively shovel grain into a chute to trigger events like feeding a child or cleaning up a kitchen.

*Space Octopus Mono* by Matthew Gatland: An 8-bit style arcade game where the players control the horizontal position of the spaceship via wooden sliders on wooden rails.

*Off Da Railz* by Vox Populi: A game where the player controls a train with a wooden board that has tilt sensors for direction and speed control.

*CatManDudu* by Emile Drescher and Tom Tyer-Drake: An experimental game controlled by two foot-operated buttons for direction and a toilet chain switch for triggering "shots".

*Word Wars* by Jenna Gavin and Tom Tyer-Drake: A competitive game in which up to eight players form words by "grabbing" letters that appear on the screen by pushing a single button.

*Fruit Racers* by Jenna Gavin and Tom Tyer-Drake: A four player competitive game with rotary encoders as input devices to control the direction of fruit on the screen in a race setting.

The event was visited by more than 100 people, and also featured on an evening TV show [27]. Figure 1 shows gamers at the arcade interacting with a number of the games.

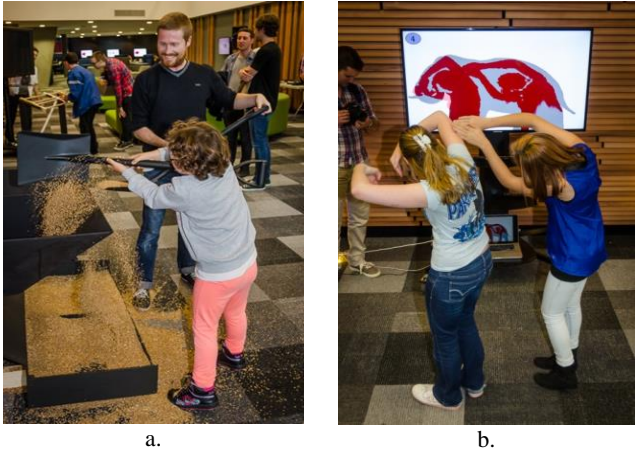

a.　　　　　　　　　b.
**Figure 1. Gameplay at the Rabble Room Arcade.**

Figure 2 and Figure 3 highlight two of the interfaces developed for the games. Figure 2 shows the rotary encoders used in the game "Fruit Racers" whereas Figure 3 shows the ultrasonic sensors and sliding rails used in "Space Octopus Mono".

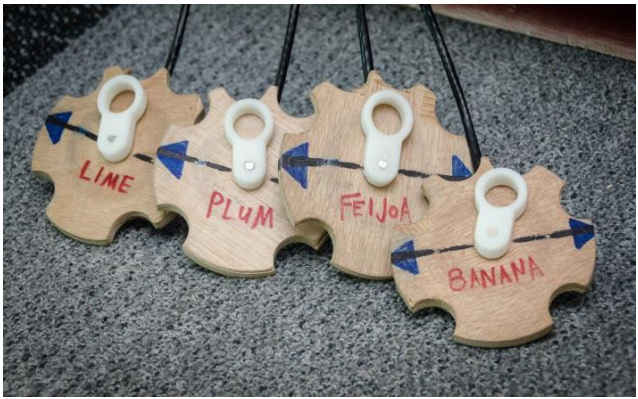

**Figure 2.** Fruit Racers.

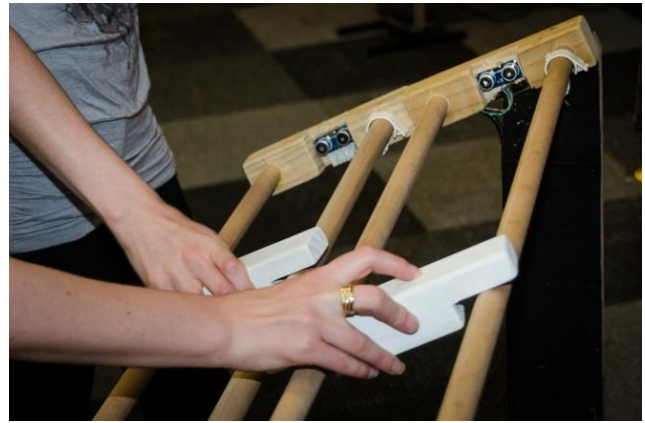

**Figure 3.** Space Octopus Mono.

Of particular interest is the game "Word Wars", conceived in 2013. The game springs from a minimalist design perspective, namely, the investigation of game mechanics that arise if each player only has one button. It has a simple ruleset, is casual yet tense, and encourages a tight social experience around a waist-height cabinet. It is built in Processing for the purpose of receiving multiple button presses through Arduino, where a standard keyboard will limit simultaneous key presses to six. Gameplay in Word Wars is based around completing an English word more than three letters long. Each player pushes their button to try and grab the letter in the middle of the shared screen. In its own way, this game challenges the pervading view that the "button" is impeding game media development and rejects the call to "discard the button in favour of natural interfaces". It clearly demonstrates that the simple button can in fact facilitate more social game play and be used in innovative and exciting game designs. This suggests that perhaps the humble button is not a major issue, but instead the lack of creativity in designing play in a fun and engaging manner.

## 4. OBSERVATIONS

The key theme for reflecting on the successes of the event relate simply to the idea of engagement. In this context, engagement can be considered at three different levels, engagement with "the idea of the event", engagement with the event itself and engagement with the games and game play. Engagement with the idea of the event cuts across multiple facets, and the first interesting point to consider is the number of games contributed by game studios or individuals outside of the event team. Of the eight games showcased, only two games were fully developed by the event team. The interfaces for the remaining six were developed by the event team, however in most cases the games themselves were developed either wholly by or in conjunction with external contributors. Given the relatively short timescale for development, this suggests that the gaming community is inherently social and is looking for opportunities to engage in unique, sociable play spaces. It is encouraging that the spirit of the anarchic arcades of the 1970s is still present in the game development community.

Evidence of engagement with the event was also positive, with over 100 attendees all of whom embraced the alternative interfaces and clearly identified with the makeshift and local spirit of the event. In terms of the engagement with the games themselves, it was clear that the interfaces promoted a more active gameplay and greater engagement between the players, as well as between the players and the games. All of the interfaces were "inefficient by design" and there exists an opportunity to investigate the value of such interfaces. Another consideration for future work is to explore how the interfaces facilitate both cooperation and competition between players. Exploring both of these areas will provide more formal evidence that the interfaces increase engagement.

## 5. CONCLUSION

Tangible interfaces have had a long history in video gaming, especially in the mechanical cabinets and the arcade machines pre-1990. New technology has enabled the layperson to create functional prototypes with ease, using these developments to explore unique, independent, and physical video games. Community based play events can embrace the physicality of play as a means of increasing engagement and promoting the development of the gaming medium. Inefficiency of interface and interactions are *spectacle*, and are well-received in light-hearted social settings. The Rabble Room Arcade event demonstrated local acceptance and engagement with unusual physical interfaces. The design of "Word Wars" demonstrates meanwhile that interfaces, when used creatively, need not dismiss the button as restrictive or detached.

## 6. REFERENCES

[1] Williams, D. 2002. Structure and competition in the US home video game industry. *International Journal on Media Management*. 4, 1 (May 2002), 41-54.

[2] Voida, A. and Greenberg, S. 2009. Wii all play: the console game as a computational meeting place. In *Proceedings of the SIGCHI Conference on Human Factors in Computing Systems* (Boton, MA, USA, April 4-9, 2009). CHI '09. ACM, New York, NY, 1559-1568. DOI= http://doi.acm.org/10.1145/1518701.1518940.

[3] Escobar-Chaves, S. L. and Anderson, C. A. 2008. Media and risky behaviors. *The Future of Children*. 18, 1 (Spring 2008), 147-180.

[4] Bailey, K., West, R. and Anderson, C. A. 2010. A negative association between video game experience and proactive cognitive control. *Psychophysiology*. 47, 1 (January 2010), 34-42.

[5] Gentile, D. 2009. Pathological Video-Game Use Among Youth Ages 8 to 18 A National Study. *Psychological Science*. 20, 5 (May 2009), 594-602.

[6] Anderson, C. A. , Shibuya, A. , Ihori, N., Swing, E. L., Bushman, B. J., Sakamoto, A., Rothstein, H. R. and Saleem, M. 2010. Violent video game effects on aggression, empathy, and prosocial behavior in eastern and western countries: a meta-analytic review. *Psychological Bulletin*, 136, 2, (March 2010), 151-173.

[7] Anderson, C. A., Gentile, D. A. and Buckley, K. E. 2007. *Violent video game effects on children and adolescents*. Oxford University Press, Oxford.

[8] Lee, J., Luchini, K., Michael, B., Norris, C. and Soloway, E. 2004 "More than just fun and games: Assessing the value of educational video games in the classroom," In *CHI '04 Extended Abstracts on Human Factors in Computing Systems* (Vienna, Austria, April 24-29, 2004). CHI EA '04. ACM, New York, NY, USA, 1375-1378. DOI= http://doi.acm.org/10.1145/985921.986068.

[9] Gentile, D. A. , Anderson, C. A., Yukawa, C. A., Ihori, N., Saleem, M., Ming, L. K., Shibuya, A., Liau, A. K., Khoo, A. and Bushman, B. J. 2009. The effects of prosocial video games on prosocial behaviors: International evidence from correlational, longitudinal, and experimental studies. *Personality and Social Psychology Bulletin*. 35, 6 (March 2009), 752-763.

[10] Taylor, L. N. 2006. Positive features of video games. In Dowd, N.E, Singer, D.G. and Wilson, R.F. eds. Handbook of children, culture and violence, Sage Publications, New York, 247-265.

[11] Rhodes, R. E., Warburton, D. E. and Bredin, S. S. 2009. Predicting the effect of interactive video bikes on exercise adherence: An efficacy trial. *Psychology, health & medicine*. 14, 6, 631-640.

[12] Sell, K., Lillie, T. and Taylor, J. 2008. Energy expenditure during physically interactive video game playing in male college students with different playing experience. *Journal of American College Health*, 56, 5, 505-512.

[13] Kent, S.L. 2001. *The first quarter: A 25-year history of video games*. BWD Press, New York.

[14] Williams, D. 2005. A Brief Social History of Game Play. Retrieved August 15 2014: http://is.muni.cz/el/1421/podzim2013/IM082/um/WilliamsSocHist.pdf

[15] Wolf, M.J. 2008. *The video game explosion: a history from PONG to Playstation and beyond*. ABC-CLIO, Santa Barbara.

[16] Salen, K. and Zimmerman, E. 2004. *Rules of play: Game design fundamentals*. MIT Press, Boston.

[17] Griffin, S. 2005. Push. Play: An examination of the gameplay button. In *Proceedings of the 2005 DiGRA Conference: Changing Views – Worlds in Play* (Vancouver, June 16-20, 2005). DiGRA, Tampere.

[18] Parker, J, 2008. Buttons, simplicity, and natural interfaces. *Loading...* ,2, 2. Retrieved August 15 2014: http://journals.sfu.ca/loading/index.php/loading/article/view/33/30.

[19] Hoysniemi, J. 206. International survey on the Dance Dance Revolution game. *Computers in Entertainment*, 4, 2, Article 8 (April 2006). DOI= http://doi.acm.org/10.1145/1129006.1129019p.

[20] Jansz, J. and Martens, L. 2005. Gaming at a LAN event: the social context of playing video games. *New Media & Society*, 7, 3 (June 2005), 333-355, 2005.

[21] De Kort, Y.A. and Ijsselsteijn, W. A. 2008. People, places, and play: player experience in a socio-spatial context. *Computers in Entertainment*. 6, 2, Article 18 (April/June 2008). DOI= http://dx.doi.org/10.1145/1371216.1371221.

[22] Pasternack, A. 2011. *Oral History of Gaming: Babycastles, the DIY Arcade*. Retrieved August 15 2014: http://motherboard.vice.com/read/motherboard-tv-babycastles-the-diy-arcade

[23] McGonigal, J. 2011. *Reality is broken: Why games make us better and how they can change the world*. Penguin, London.

[24] Behm-Morawitz, E. and Mastro, D. 2009. The effects of the sexualization of female video game characters on gender stereotyping and female self-concept. *Sex Roles*, 61, 11/12 (August 2009), 808-823.

[25] Fortin, C., DiPaola, S., Hennessy, K., Bizzocchi, J. and Neustaedter, C. 2013. Medium-specific properties of urban screens: towards an ontological framework for digital public displays. In *Proceedings of the 9th ACM Conference on Creativity & Cognition* (Sydney, Australia, June 17-20, 2013). C&C '13. ACM, New York, NY, USA, 243-252. DOI= http://doi.acm.org/10.1145/2466627.246662.

[26] Wilson, D. 2011. Brutally unfair tactics totally ok now: On self-effacing games and unachievements. *Game Studies*. 11, 1 (February 2011). Retrieved August 15 2014: http://gamestudies.org/1101/articles/Wilson 2011.

[27] TVNZ. 2013. *The Future of Gaming?* Retrieved August 15 2014: http://tvnz.co.nz/seven-sharp/future-gaming-video-56240